\documentclass[
    ,final            
  ]
  {aipproc}
\usepackage{graphicx,psfrag}

\layoutstyle{8x11single}
\setcitestyle{numbers}

\begin{document}

\title{Matching the quark model to the 1/$N_c$ expansion}


\author{Dan Pirjol}{
  address={ National
Institute for Physics and Nuclear Engineering, Department of Particle
Physics, \\  077125 Bucharest, Romania}
}

\author{Carlos Schat\thanks{Speaker} }
{
  address={Department of Physics and Astronomy, Ohio University, Athens, Ohio 45701, USA \\
           Departamento de F\'{\i}sica, FCEyN,
Universidad de Buenos Aires, Ciudad Universitaria, Pab.1, (1428) Buenos
Aires, Argentina} 
}

\begin{abstract}
We compute the coefficients of the effective mass operator of
the $1/N_c$ expansion for negative parity $L=1$ 
excited baryons using the Isgur-Karl model in order to compare the general 
approach, where the coefficients are obtained by fitting to data, 
with a specific constituent quark model calculation.
We discuss the physics behind 
the fitted coefficients  
for the scalar part of the most general two-body quark-quark interaction. 
We find that both pion exchange and gluon exchange lead 
to the dominance of the 
same operator at the level of the effective mass operator, which is also observed from data.
\end{abstract}

\maketitle

\section{Introduction}

Quark models provide a simple picture of the physics of 
ground state baryons and their excitations \cite{De Rujula:1975ge,Isgur:1977ef}, 
but their connection with QCD is not so clear. 
An alternative approach to baryon phenomenology that is closer 
connected to QCD is provided by the $1/N_c$ expansion 
\cite{Dashen:1993jt}, which has been applied to the study of both the ground state and excited nucleons
\cite{Goity:1996hk,Pirjol:1997bp,Carlson:1998vx,Schat:2001xr,Goity:2003ab,Pirjol:2003ye,Matagne:2004pm}, 
leading to interesting insights that would be difficult to obtain in a particular model calculation.

In order to bridge the gap between the more intuitive quark model 
calculations and the effective operator approach, in a recent paper 
\cite{Pirjol:2007ed} we showed how to match an 
arbitrary quark model Hamiltonian onto the operators of the $1/N_c$ expansion.
We made use of the transformation of the states and operators
under the permutation group of $N_c$ objects acting
on the spin-flavor of the quarks. The approach is similar to the one 
discussed in Ref.~\cite{Collins:1998ny} for the orbital part of the interaction and at $N_c=3$. 
The main result of \cite{Pirjol:2007ed} can be summarized as follows:
consider a two-body quark Hamiltonian $V_{qq} = \sum_{i<j} O_{ij} R_{ij}$,
where $O_{ij}$ acts on the spin-flavor quark degrees of freedom and $R_{ij}$
acts on the orbital degrees of freedom. Then the hadronic matrix elements of
the quark Hamiltonian on a baryon state $|B\rangle$ contains only the projections
$O_\alpha$  onto irreducible representations of the spin-flavor permutations
and can be factorized as $\langle B |V_{qq}|B\rangle = \sum_\alpha C_\alpha 
\langle O_\alpha\rangle$, where the  coefficients $C_\alpha$ are 
reduced matrix elements of the orbital operators $R_{ij}$, 
given by overlap integrals of the quark model wave functions.

The matching procedure has been discussed in detail 
for the Isgur-Karl (IK) model in Ref.~\cite{Galeta:2009pn}.
As an application of the general approach
in Ref.~\cite{Pirjol:2008gd,Pirjol:2010th} we 
examine the spectrum of  $L=1$ negative parity baryons and discuss the 
implications for the spin-flavor structure of 
the most general two-body quark interaction. 
The present  talk
summarizes some  aspects of this recent work that strive to 
understand
the physics hidden in the numerical values of the 
fitted coefficients of the general  operator expansion. 

\section{The mass operator of the Isgur-Karl model}
\label{IKV}

The Isgur-Karl model is defined by the quark Hamiltonian
\begin{eqnarray}
{\cal H}_{IK} = {\cal H}_0 + {\cal H}_{\rm hyp}  \,, 
\end{eqnarray}
where
${\cal H}_0$ contains the confining potential and kinetic terms of the quark
fields, and is symmetric under spin and isospin. The hyperfine
interaction ${\cal H}_{\rm hyp}$ is given by 
\begin{eqnarray}\label{HIK} 
{\cal H}_{\rm hyp} = A \sum_{i<j}\Big[ \frac{8\pi}{3} \vec s_i \cdot \vec s_j
\delta^{(3)}(\vec r_{ij}) + \frac{1}{r_{ij}^3} (3\vec s_i \cdot \hat r_{ij} \
\vec s_j \cdot \hat r_{ij} - \vec s_i\cdot \vec s_j) \Big] \,, 
\end{eqnarray} 
where $A$ determines the strength of the interaction, and
$\vec r_{ij} = \vec r_i - \vec r_j$ is the distance between quarks
$i,j$.  The first term is a local spin-spin interaction, and the second
describes a tensor interaction between two dipoles. This interaction
Hamiltonian is an approximation to the gluon-exchange interaction,
neglecting the spin-orbit terms.

 In the original formulation of the 
IK model \cite{Isgur:1977ef} the confining forces are harmonic.  
We will derive in the following the form 
of the mass operator without making any assumption on the shape of the confining quark forces. 
We refer to this more general version of the model as IK-V(r).

The $L=1$ quark model states include the following SU(3) multiplets: 
two spin-1/2 octets $8_\frac12, 8'_\frac12$, two spin-3/2 octets $8_\frac32, 8'_\frac32$,
one spin-5/2 octet $8'_\frac52$, two decuplets $10_\frac12, 10_\frac32$ and two singlets
$1_\frac12, 1_\frac32$. States with the same quantum numbers mix. For the 
$J=1/2$ states we define
the relevant mixing angle $\theta_{N1}$ 
in the nonstrange sector as
$
N(1535)  =   \cos\theta_{N1} N_{1/2} + \sin\theta_{N1} N'_{1/2} \ , \
N(1650)  =   -\sin\theta_{N1} N_{1/2} + \cos\theta_{N1} N'_{1/2} 
$
and similar equations for the $J=3/2$ states, which define a second mixing angle 
$\theta_{N3}$.

We find \cite{Galeta:2009pn} that the most general
mass operator in the IK-V(r)  model depends only on three unknown orbital overlap
integrals, plus an additive constant $c_0$ related to the matrix element
of ${\cal H}_0$, and can be written as 
\begin{eqnarray}\label{IKMass} 
\hat M =
c_0 \mathbf{1} + a  S_c^2 + b L_2^{ab} \{ S_c^a\,, S_c^b\}  + c L_2^{ab} \{
s_1^a\,, S_c^b\} \,,
\end{eqnarray} 
where the spin-flavor operators are
understood to act on the state $|\Phi(SI)\rangle$ constructed as a
tensor product of the core of quarks 2,3 and the `excited' quark 1, as given in 
\cite{Carlson:1998vx,Pirjol:2007ed}.
  The
coefficients are given by
$
a = \frac12 \langle R_S\rangle \ , \ 
b = \frac{1}{12} \langle Q_S\rangle - \frac16 \langle Q_{MS}\rangle \ , \  
c = \frac16 \langle Q_S\rangle + \frac16 \langle Q_{MS}\rangle \ .
$
The reduced matrix elements $R_S,Q_S,Q_{MS}$ for the orbital part of the interaction contain 
the unknown spatial dependence and are defined in Refs.~\cite{Carlson:1998vx,Galeta:2009pn}.
Evaluating the matrix elements using Tables~II,~III in
Ref.~\cite{Carlson:1998vx} we find the following explicit result for
the mass matrix 
\begin{equation}
\begin{array}{cc}
M_{1/2} = 
\left( 
\begin{array}{cc}
c_0 + a                   &  -\frac53 b + \frac{5}{6}c \\ 
-\frac53 b + \frac{5}{6}c & c_0 + 2a + \frac53(b+c)\\ 
\end{array} 
\right) \,, & 
M_{3/2} = 
\left(
\begin{array}
{cc} c_0 + a                                  &  \frac{\sqrt{10}}{6} b -\frac{\sqrt{10}}{12}c \\ 
\frac{\sqrt{10}}{6} b - \frac{\sqrt{10}}{12}c & c_0 + 2a - \frac43(b+c)\\ 
\end{array}
\right) \,, 
\end{array} 
\end{equation}
and $M_{5/2}      = c_0 + 2a +\frac13 (b+c) \,, \Delta_{1/2} = \Delta_{3/2} = c_0 + 2a $ .
Computing the reduced matrix elements with the interaction given by Eq.~(\ref{HIK}),  one finds that the
reduced matrix elements  in the IK model with harmonic oscillator
wave functions are all related and can be expressed in terms of the single parameter $\delta$ as
$
\langle Q_{MS}\rangle = \langle Q_S\rangle = -
\frac35 \delta \ ; \  \langle R_S\rangle = \delta \ .
$
This gives a relation among
the coefficients $a,b,c$ of the mass matrix Eq.~(\ref{IKMass}): 
$
a = \frac12 \delta \,, b = \frac{1}{20} \delta \,,
c = - \frac15 \delta \,.
$
We recover the well known result that in the harmonic oscillator model,  the entire spectroscopy of 
the $L=1$ baryons is fixed by one
single constant $\delta=M_\Delta-M_N \sim 300 \  {\rm MeV}$, along with an overall additive constant $c_0$, and
the model becomes very predictive.
\begin{figure}[t!]
\includegraphics[width=7.0cm]{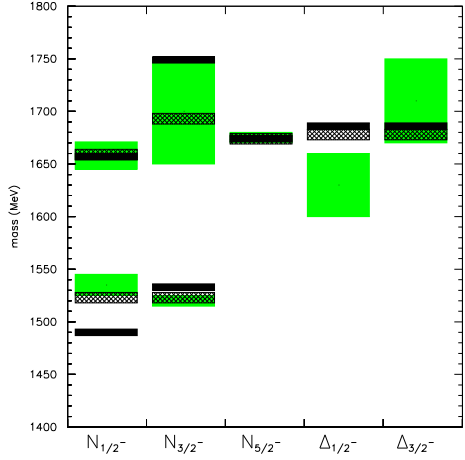}
\caption{Masses predicted by the IK model (black bars), by the IK-V(r) model (hatched bars) 
and the experimental masses (green boxes) from Ref.~\cite{Amsler:2008zzb}. }
\label{fig:masses}
\end{figure}
In Fig.~\ref{fig:masses} we show the result of a best fit of $a,b,c$ in the IK-V(r) model together 
with the predictions of the IK model. The IK-V(r) spectrum is the best fit possible for a 
potential model with the spin-flavor interaction given in Eq.~(\ref{HIK}).

\section{Testing a more general Spin-Flavor structure}
\label{Sec:Hamiltonian}

The most general quark Hamiltonian containing only two-body interactions
has the form \cite{Pirjol:2008gd}
\begin{eqnarray}\label{1}
H_{qq} = H_0 + \sum_{i<j} (f_1(\vec r_{ij})  t_i^a t_j^a + 
f_2(\vec r_{ij})  \vec s_i \cdot \vec s_j + 
f_3(\vec r_{ij})  \vec s_i t_i^a \cdot \vec s_j t_j^a)  + H_{s-o} + H_q \, ,
\end{eqnarray}
where 
$H_0$ is the part of the quark Hamiltonian which does not depend on the quarks spin
and flavor degrees of freedom. We show explicitly only the part of the Hamiltonian
which transforms as a scalar ($\ell=0$) under $SO(3)$, the group of orbital rotations - the scalar part 
of the quark Hamiltonian. The $H_{s-o}, H_q $ denote the spin-orbit and the quadrupole interaction,  
which transform as a vector ($\ell=1$) and a 
tensor ($\ell=2$), 
respectively.

We will consider  the case of  contact scalar interactions
\begin{eqnarray}
f_\nu (\vec r_{ij}) = A_\nu \delta^{(3)}(\vec r_{ij})\,,\qquad \nu=1,2,3,
\end{eqnarray}
and study the following question: what information can be obtained from the coefficients
$c_k$ of the $1/N_c$ studies of the spectrum of $L=1$ negative parity baryons? 

The quark Hamiltonian $H_{qq}$ is matched onto the hadronic mass
operator
\begin{eqnarray}\label{scalar}
M = c_0 \mathbf{1} + c_1T^2 + c_2 S_c^2 + c_3 \vec s_1 \cdot \vec S_c + \cdots
\end{eqnarray}
where the ellipses denote terms arising from the tensor and spin-orbit interactions, which 
were considered in Ref.~\cite{Pirjol:2008gd}. 
All three possible zero range two-body interactions  $O_{ij}^{(\nu)} = t_i^a t_j^a,
\vec s_i \cdot \vec s_j $ and $  \vec s_i t_i^a \cdot \vec s_j t_j^a$ lead to $c_1=c_3=0$
and
are matched onto the
same operator $O_2 = S_c^2$ in the effective theory. This means that there
is  no way to distinguish between these three types of scalar interactions as 
long as  they are
of very short range. 
Experimentally, at $N_c=3$ one can determine
only two linear combinations of the three coefficients $c_{1,2,3}$ (as functions of
$\theta_{N1,N3}$) \cite{Pirjol:2008gd} from the mass spectrum and 
mixing angles of the negative parity $L=1$ baryons. These linear combinations can be taken as
$ \tilde c_1 = c_1 - \frac12 c_3 \,, 
 \tilde c_2 = c_2 + c_3\,
$
and their explicit expressions in terms of the nonstrange hadron masses
and mixing angles can be found in Ref.~\cite{Pirjol:2010th}.

\begin{figure}[t!]
\psfrag{XXX}[][][1.5]{$\tilde c_1/\tilde c_2$}
\psfrag{T}[t][][1.5]{$\theta_{N1}$}
\includegraphics[width=6.0cm]{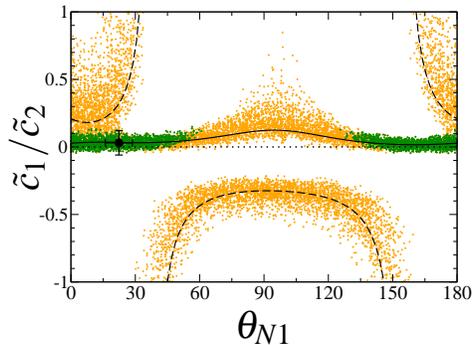}
\caption{Scatter plot for the ratio of
coefficients $\tilde c_1/\tilde c_2$ as a function of the mixing angle $\theta_{N1}$.}
\label{fig:ratioc12}
\end{figure}

A first estimate of  $\tilde c_{1,2}$ can be made using the mixing angles $\theta_{N1, N3}$ determined from a fit to
$N^*$ strong decays and photoproduction data 
$(\theta_{N1}, \theta_{N3}) = (0.39 \pm 0.11, 2.82 \pm 0.11) = (22^\circ \pm 6^\circ,
162^\circ \pm 6^\circ)$ \cite{Goity:2004ss,Scoccola:2007sn}.
Using these values and the  
hadron masses from the PDG \cite{Amsler:2008zzb} we obtain
$
\tilde c_1 = 3.9 \pm 11.0 \mbox{ MeV} \,,
\tilde c_2 = 129 \pm 18 \mbox{ MeV}\,.
$
The corresponding ratio of the coefficients $\tilde c_i$ displays the relative suppression of $\tilde c_1$
\begin{eqnarray}\label{cratio}
\tilde c_1/\tilde c_2 = 0.03 \pm 0.09\,. 
\end{eqnarray}
An alternative determination of this ratio can be made using only the excited baryon masses, 
as discussed in Ref.~\cite{Pirjol:2008gd}. In this paper it was shown that in 
any quark model containing only two-body quark interactions there is a correlation 
between both  mixing angles
(up to a discrete ambiguity) that corresponds to the dashed and solid lines in Fig.~\ref{fig:ratioc12}.
The scatter plot takes into account the experimental errors in the 
determination of the baryon masses.
A second  relation (Eq.~(6) in Ref.~\cite{Pirjol:2008gd}) 
expressing the spin-average 
of the $SU(3)$ singlet states $\bar\Lambda = \frac13 \Lambda_{1/2} + \frac23 \Lambda_{3/2}$ 
in terms of the nonstrange states further constrains the mixing angle and ratio  
to the dark (green) area in the plot and singles out the solid line as the preferred solution.
In Fig.~\ref{fig:ratioc12} the first determination of the ratio Eq.~(\ref{cratio}) 
is shown as the black point with error bars and falls within the dark (green) area.

Finally, it is interesting to notice \cite{Carlson:1998vx,Pirjol:2010th}
 that for the particular case ($\nu=3$) of the spin-flavor structure 
corresponding to 
one-pion exchange  
$\tilde c_1$ vanishes regardless of the spatial range of the orbital part of the 
interaction 
(although for non-zero range  $c_{1,3}$ do not vanish separately).

\section{Conclusions}

We presented the effective mass operator of the Isgur-Karl model in the operator basis 
used in
the $1/N_c$ expansion for excited baryons. This explicit model calculation of the operator 
coefficients  (for details 
see Ref.~\cite{Galeta:2009pn}) connects 
two different approaches to baryon phenomenology and is
an illustration of the general matching procedure discussed in Ref.~\cite{Pirjol:2007ed}
using the permutation group.

Considering the most general spin-flavor structure for the scalar 
part of the  two-body quark interactions we find
\cite{Pirjol:2010th}
 that in the case of
contact interactions all spin-flavor structures are matched onto the 
same effective operator in the $1/N_c$ expansion. For the particular 
case of a pion exchange interaction this holds independently of the 
range of the spatial part of the 
interaction. Comparing the values obtained for the operator 
coefficients from masses and independent determinations of mixing angles we 
find that a mixture of pion exchange and massless gluon exchange interactions 
is compatible with data, giving support to the physical picture 
considered in 
Refs.~\cite{Manohar:1983md,Weinberg:2010bq}.

\begin{theacknowledgments}
The work of C.S. was supported by CONICET and partially supported by  the U.~S.  Department of Energy, Office of
Nuclear Physics under contract No. DE-FG02-93ER40756
with Ohio University.
\end{theacknowledgments}


\begin{thebibliography}{99}


\bibitem{De Rujula:1975ge}
  A.~De Rujula, H.~Georgi and S.~L.~Glashow,
  Phys.\ Rev.\  D {\bf 12}, 147 (1975).

\bibitem{Isgur:1977ef}
  N.~Isgur and G.~Karl,
  Phys.\ Lett.\  B {\bf 72}, 109 (1977).

\bibitem{Dashen:1993jt}
  R.~F.~Dashen, E.~Jenkins and A.~V.~Manohar,
  Phys.\ Rev.\  D {\bf 49}, 4713 (1994)
  [Erratum-ibid.\  D {\bf 51}, 2489 (1995)];
R.~F.~Dashen, E.~Jenkins and A.~V.~Manohar,
  Phys.\ Rev.\  D {\bf 51}, 3697 (1995).

\bibitem{Goity:1996hk}
  J.~L.~Goity,
  Phys.\ Lett.\  B {\bf 414}, 140 (1997).

\bibitem{Pirjol:1997bp}
  D.~Pirjol and T.~M.~Yan,
  Phys.\ Rev.\ D {\bf 57}, 1449 (1998);
  Phys.\ Rev.\ D {\bf 57}, 5434 (1998).

  
  
\bibitem{Carlson:1998vx}
C.~E.~Carlson, C.~D.~Carone, J.~L.~Goity and R.~F.~Lebed,
  Phys.\ Lett.\  B {\bf 438}, 327 (1998);
  Phys.\ Rev.\  D {\bf 59}, 114008 (1999).

\bibitem{Schat:2001xr}
  C.~L.~Schat, J.~L.~Goity and N.~N.~Scoccola,
  Phys.\ Rev.\ Lett.\  {\bf 88}, 102002 (2002);
J.~L.~Goity, C.~L.~Schat and N.~N.~Scoccola,
  Phys.\ Rev.\  D {\bf 66}, 114014 (2002).

\bibitem{Goity:2003ab}
  J.~L.~Goity, C.~Schat and N.~N.~Scoccola,
  Phys.\ Lett.\  B {\bf 564}, 83 (2003).

\bibitem{Pirjol:2003ye}
  D.~Pirjol and C.~Schat,
  Phys.\ Rev.\ D {\bf 67}, 096009 (2003);
  AIP Conf.\ Proc.\  {\bf 698}, 548 (2004).

\bibitem{Matagne:2004pm}
  N.~Matagne and F.~Stancu,
  Phys.\ Rev.\  D {\bf 71}, 014010 (2005);
  Phys.\ Lett.\  B {\bf 631}, 7 (2005);
  Phys.\ Rev.\  D {\bf 74}, 034014 (2006).


\bibitem{Pirjol:2007ed}
  D.~Pirjol and C.~Schat,
  Phys.\ Rev.\  D {\bf 78}, 034026 (2008).

\bibitem{Collins:1998ny}
  H.~Collins and H.~Georgi,
  Phys.\ Rev.\  D {\bf 59}, 094010 (1999).

\bibitem{Galeta:2009pn}
  L.~Galeta, D.~Pirjol and C.~Schat,
  Phys.\ Rev.\  D {\bf 80}, 116004 (2009).

\bibitem{Pirjol:2008gd}
  D.~Pirjol and C.~Schat,
Phys.\ Rev.\ Lett.\ {\bf 102}, 152002 (2009).


\bibitem{Pirjol:2010th}
  D.~Pirjol and C.~Schat,
  arXiv:1007.0964 [hep-ph].


\bibitem{Goity:2004ss}
J.~L.~Goity, C.~Schat and N.~Scoccola,
 Phys.\ Rev.\  D {\bf 71}, 034016 (2005).

\bibitem{Scoccola:2007sn}
  N.~N.~Scoccola, J.~L.~Goity and N.~Matagne,
  Phys.\ Lett.\  B {\bf 663}, 222 (2008).

  \bibitem{Amsler:2008zzb}
  C.~Amsler {\it et al.}  [Particle Data Group],
  Phys.\ Lett.\  B {\bf 667}, 1 (2008).

\bibitem{Manohar:1983md}
  A.~Manohar and H.~Georgi,
  Nucl.\ Phys.\  B {\bf 234}, 189 (1984).

\bibitem{Weinberg:2010bq}
  S.~Weinberg,
  arXiv:1009.1537 [hep-ph].


\end{thebibliography}
\end{document}